\documentclass{elsart}

\usepackage{graphicx}

\usepackage{amssymb,amsmath}
\newcommand{\omegagw}{\Omega_{\rm \scriptscriptstyle GW}}

\begin{document}

\begin{frontmatter}


\title{Evolution of gravitational waves in the 
high-energy regime of brane-world cosmology}
%
\author{Takashi Hiramatsu\corauthref{utap}},
\ead{hiramatsu@utap.phys.s.u-tokyo.ac.jp}
\author{Kazuya Koyama\corauthref{utap}\corauthref{port}},
\ead{kazuya@utap.phys.s.u-tokyo.ac.jp,Kazuya.Koyama@port.ac.uk}
\author{Atsushi Taruya\corauthref{utap}\corauthref{resceu}}
\ead{ataruya@utap.phys.s.u-tokyo.ac.jp}
\address[utap]{Department of Physics , University of Tokyo, Tokyo 113-0033, Japan}
\address[port]{Institute of Cosmology and Gravitation, University of
 Portsmouth, Portsmouth PO1 2EG, UK}
\address[resceu]{Research Center for the Early Universe, University of Tokyo, Tokyo 
113-0033, Japan}
%
%
%
%
%
\begin{abstract}
We discuss the cosmological evolution of gravitational 
waves (GWs) after inflation in a brane-world cosmology embedded in 
five-dimensional anti-de Sitter (AdS${}_5$) bulk spacetime. 
In a brane-world scenario, the evolution of GWs is affected by the 
non-standard cosmological expansion and the excitation of the
Kaluza-Klein modes (KK-modes), which are significant in the 
high-energy regime of the universe. 
We numerically solve the wave equation of GWs in the  
Poincar\'e coordinates of the AdS${}_5$ spacetime. 
Using a plausible initial condition from inflation, 
we find that, while the behavior of GWs in the bulk is 
sensitive to the transition time from inflation to the 
radiation dominated epoch, the amplitude of GWs on 
the brane is insensitive to this time if the transition occurs 
early enough before horizon re-entry. 
As a result, the amplitude of GWs is suppressed by 
the excitation of KK-modes which escape from the brane 
into the bulk, and the effect may compensate the enhancement 
of the GWs by the non-standard cosmological expansion. 
Based on this, the influence of the high-energy effects on the 
stochastic background of GWs is discussed. 
\end{abstract}
%
%
%
%
%
%
%
%
%
%
\end{frontmatter}
%
%

\section{Introduction}
\label{sec: introduction}
%
%
%
The stochastic background of gravitational waves (SBGW) generated 
during the accelerating phase of the early universe is one of the 
most fundamental predictions of the inflationary scenario and this 
can provide a direct way to probe the very early universe. 
In particular, information about extra-dimensions is 
expected to be imprinted on the SBGW. Motivated by recent developments in
particle physics, the possibility that our universe is described by  
brane with three spatial dimensions embedded in a higher-dimensional space has been
extensively discussed. According to this scenario, gravity propagates in
the extra spatial dimensions, while the standard model particles are
confined to the three dimensional brane. At low-energy scales, four-dimensional
general relativity is successfully recovered and the extra-dimensional
effects should be fairly small. On the other hand, at high-energy
scales, the localization of gravity is not always guaranteed and a
significant deviation from the standard four-dimensional theory is
expected. If this is true, the SBGW can provide a direct probe of the
extra-dimensional effects.

In this letter, we focus on the brane-world model proposed by Randall
and Sundrum \cite{RS2}. In this model, a three dimensional brane is
embedded in the five
dimensional anti-de Sitter (AdS${}_5$) bulk spacetime (see Ref.\cite{Maartens}
for a comprehensive review of this brane-world scenario). The important
model parameter is the AdS${}_5$ curvature radius $\ell$ that determines
the scale on which Newton's force law is modified. 
Table-top experiments on Newton's force law impose the
constraint on $\ell$ that $\ell<0.1$mm \cite{Chi}. 
In the context of cosmology, the evolution of the universe is 
significantly modified by extra-dimensional effects when the Hubble 
horizon becomes shorter than the AdS${}_5$ curvature scale, $H \ell >1$. 
Thus there is  a critical frequency 
corresponding to the wavelength of the gravitational waves (GWs) 
that cross the horizon when $H \ell =1$. This frequency is given by
$f_{\rm crit}=0.2$mHz$\,(0.1$mm$/\ell)^{1/2}$ \cite{Hogan}.
Since the GWs with short wavelengths re-enter the horizon 
at high energies, it is expected that the SBGW above the critical
frequency is crucially affected by the extra-dimensional effects. 
Theoretically, there are two main effects in the high-energy 
regime: (i) the cosmological expansion becomes slow compared with that 
in four-dimensional general relativity, which enhances 
the amplitude of the SBGW and 
(ii) the excitation of Kaluza-Klein modes (KK-modes) 
suppresses the amplitude of the GWs on the brane. 
An interesting point is that these two effects affect the 
amplitude of the SBGW in an opposite way. Thus, quantitative
calculations of the effects of the KK-modes excitation are 
essential in predicting the spectrum of SBGW. 

There have been many attempts to understand the effect of 
the KK-mode excitation. One way is to work with Gaussian normal
coordinates \cite{ELMW,Battye1}.
Using these coordinates, we have performed a numerical study of 
the high-energy effects and showed that 
the excitation of KK-modes dominates over the effect of the non-standard
cosmic expansion \cite{TH}. However, the Gaussian normal coordinates have 
a coordinate singularity and the analysis is limited to 
the relatively low-energy universe $(H \ell <1)$. 
Recently, Ichiki and Nakamura  have performed 
a numerical calculation in the high-energy regime $(H \ell >1)$ 
using a null coordinate system \cite{IN1,IN2}. 
While they reached the same 
conclusion as our previous study, they took the initial 
condition so that the perturbation is constant along a null
hypersurface. At present, however, it remains unclear whether 
the initial condition imposed by them is appropriate for 
the one determined from inflation.

In this letter, we use the Poincar\'e coordinate system to perform numerical 
calculations. The main advantage of this coordinate system is that 
it covers the whole bulk spacetime. Using these coordinates, several 
analytic and semi-analytic methods ahve been proposed \cite{Tanaka,KK,Battye2}, 
which work well in the low-energy regime. 
Also, the quantum fluctuations in the inflationary epoch have
been discussed in the Poincar\'e coordinate system \cite{Gor,Koba}. 
Thus, we can set more plausible initial conditions from 
the inflationary universe than those in the Gaussian normal coordinates.

We set up the basic equations and prepare the numerical calculation in
Section 2. The numerical results are presented in Section 3. We
first discuss the validity of the initial conditions. Then we attempt to 
construct the spectrum of SBGW in the high
energy regime of the universe. Finally, Section 4 is devoted to summary
and discussion.

\section{Basic equations and Numerical method}
\label{sec: basic_equation}

%
\subsection{ Background metric and evolution equation}
%

In the Randall-Sundrum single-brane model \cite{RS2}, 
a three-dimensional brane is embedded in five dimensional anti-de 
Sitter spacetime (AdS${}_5$ bulk). In this Letter, we specifically 
consider the AdS${}_5$ bulk without 
a black hole mass and assume that the matter content on the brane is simply 
described by a homogeneous and isotropic perfect fluid whose 
equation of state satisfies $p=w\rho$.

In our previous study \cite{TH}, the Gaussian normal (GN) coordinate 
system 
\begin{equation}
  ds^2 = -n^2(t,y)dt^2 + a^2(t,y)d\mathbf{x}^2 + dy^2,
  \label{eq: back_metric_GN}
\end{equation}
was used to solve the wave equation of GWs. 
In this coordinate system, the brane is located at a fixed position $y=0$. 
The lapse function $n(t,y)$ is related to the warp factor $a(t,y)$
through the relation, $n(t,y) = \dot{a}(t,y)/\dot{a}(t,y=0)$: 
\begin{eqnarray}
a(t,y)&=& a_0(t) \left\{\cosh\left(\frac{y}{\ell}\right) - 
\left(1+\frac{\rho(t)}{\lambda}\right)
\sinh\left(\frac{y}{\ell}\right)\right\},
\label{eq:warp_factor}\\
n(t,y)&=& e^{-y/\ell}+(2+3w)\,\frac{\rho}{\lambda}\,
\sinh\left(\frac{y}{\ell}\right),
\label{eq:lapse_func}
\end{eqnarray}
where $\ell$ is curvature scale of the AdS${}_5$ bulk and $\lambda>0$ is the 
tension of the brane. The quantity $a_0(t)\equiv a(t,y=0)$ denotes the 
scale factor on brane.

Although the GN coordinates (\ref{eq: back_metric_GN}) are well-behaved near 
the brane and so it is convenient to impose the boundary condition on 
the brane, difficulties arise when 
investigating the behavior of GWs in the bulk due to the coordinate 
singularity at $y=y_{\rm h}(t)$, where 
$a(t,y_{\rm h})=0$ (Eq.(\ref{eq:warp_factor})). This   
corresponds to the past null infinity of the AdS${}_5$ spacetime. 
Furthermore, a space-like $t=$const. hypersurface
approaches null near the coordinate singularity. Hence, 
a sophisticated treatment of boundary conditions near the 
singularity is required. 
As a result, the previous numerical investigation was restricted to 
the analysis at relatively low-energy scales.

In this paper, in order to extend our previous 
study to the analysis in the high-energy regime, we use the 
Poincar\'e coordinate system $(\tau,\mathbf{x},z)$. 
In the Poincar\'e coordinate system, 
the brane is moving in the static AdS${}_5$ bulk \cite{Kraus}. 
The metric is given by 
\begin{equation}
  ds^2 = \left(\frac{\ell}{z}\right)^2\{-d\tau^2 + 
(\delta_{ij}+ h_{ij})dx^idx^j+ dz^2\}, 
\label{eq: perturbed_metric}
\end{equation}
where $h_{ij}$ is the 
tensor perturbation satisfying the transverse and the traceless conditions. 
In this metric, the trajectory of the brane is described as 
$(\tau_{\rm b},\,z_{\rm b})$: 
\begin{equation}
 \tau_{\rm b} = T(t),\qquad z_{\rm b} = \frac{\ell}{a_0(t)}, 
\label{eq: branetrajectory} 
\end{equation}
where the variable $t$ is the cosmic time on brane, which has
the same meaning of time $t$ as used in the GN metric
(\ref{eq: back_metric_GN}).
From the junction conditions, the Friedmann equation and 
the conservation law become \cite{Lan}: 
\begin{equation}
 H^2(t) = \left(\frac{\dot{a_0}}{a_0}\right)^2 = \frac{\kappa_4^2}{3}
\rho\left(1+\frac{\rho}{2\lambda}\right),
\qquad 
\dot{\rho} = -3(1+w)H\rho, 
\quad
(\kappa_4^2=8\pi G),
\label{eq: Friedmann}
\end{equation}
where the dot denotes a derivative with respect to $t$ and 
$H$ is the Hubble parameter defined by $H\equiv\dot{a_0}/a_0$. 
The function $T(t)$ is given by 
\begin{equation}
 \dot{T}(t) = \frac{1}{a_0}\sqrt{1+(H\ell)^2}. \label{eq: timefunc} 
\end{equation}

Hereafter, unless otherwise mentioned, we focus on the evolution of GWs during 
the radiation dominated epoch after inflation, 
i.e., $w=1/3$. In this case, the solutions for the scale factor $a_0(t)$ and
the normalized energy density defined by 
$\epsilon(t)\equiv\rho(t)/\lambda$ are expressed as (e.g., \cite{Bin}):
\begin{equation}
 a_0(t) = a_*\left(\frac{2t^2+t\ell}{2t_*^2+t_*\ell}\right)^{1/4},
\quad 
\epsilon(t) \equiv \frac{\rho(t)}{\lambda}=\frac{\ell^2}{8t^2+4t\ell}.
\label{eq: scale_energy}
\end{equation}
The variables $a_*$ and $t_*$ are numerical constants, whose 
meanings will be given in next section.

Next consider the tensor perturbation $h_{ij}$  in the metric 
(\ref{eq: perturbed_metric}). For convenience, 
we decompose the quantity $h_{ij}$ in spatial Fourier modes as 
$h_{ij}=h_k(\tau,z) 
e^{i\mathbf{k}\cdot\mathbf{x}}\hat{e}_{ij}$, where $\hat{e}_{ij}$ 
represents a transverse-traceless polarization tensor. 
Then the evolution equation for the perturbation 
$h_k(\tau,z)$ becomes 
\begin{equation}
 \frac{\partial^2 h}{\partial \tau^2} -\frac{\partial^2 h}{\partial z^2} 
+ \frac{3}{z}\frac{\partial h}{\partial z} +k^2h= 0. 
\label{eq: wave}
\end{equation}
Here we simply omit the subscript $k$. It is known that 
the above equation has the following general solution 
(e.g., \cite{KK,Gor}): 
\begin{equation}
h(\tau,z)=\int_0^{\infty} dm\,\,\left\{
\tilde{h}_1(m)\,z^2\,H_2^{(1)}(mz)\,e^{i\,\omega\,\tau} +
\tilde{h}_2(m)\,z^2\,H_2^{(2)}(mz)\,e^{-i\,\omega\,\tau} \right\},
\label{eq:general_solution} 
\end{equation}
where $\omega^2=m^2+k^2$. The functions $H_2^{(1)}$ and $H_2^{(2)}$ 
 denote the Hankel functions of first and second kind respectively and  
the coefficients $\tilde{h}_{1,2}(m)$ are arbitrary functions of
$m$. The above expression implies that the GWs 
propagating in the bulk are generally described 
as a superposition of the zero mode ($m=0$) and the KK-modes ($m>0$). 
In the brane-world where the AdS${}_5$ bulk is bounded by the 
brane, the evolution equation of GWs must be solved by imposing
the boundary condition at the brane. 
The boundary condition at the brane is determined from the junction 
condition \cite{Maartens}. Imposing $Z_2$ symmetry on the brane, it 
is given by
\begin{equation}
 \left.\left(\frac{\partial}{\partial \tau} 
- \frac{\sqrt{1+\left(H\ell\right)^2}}{H\ell}
\frac{\partial}{\partial z}\right)h\right|_{z= z_{\rm b}(t)} = 0. 
\label{eq: junction}
\end{equation}

\subsection{Initial condition}
\label{subsec:Initial_condition}

The initial condition for the perturbed quantity $h$ 
just after inflation is determined 
by the quantum fluctuations in the inflationary epoch. 
According to Ref. \cite{Lan}, the GN coordinates (\ref{eq: back_metric_GN}) 
are a useful spatial 
slicing in the inflationary epoch and the KK-modes defined in this slicing 
are shown to be highly suppressed during the inflation. Thus, 
the zero-mode solution in the GN coordinates gives a dominant 
contribution to the metric fluctuation which is given by
\begin{equation}
h(t,y)=C (-k \eta)^{3/2} H_{3/2}^{(1)}(-k \eta),
\label{eq:inflation_zero_mode}
\end{equation}
where $C$ is a normalization constant and $\eta$ is the conformal 
time. On the other hand, the mode function given in Poincar\'e
coordinates (\ref{eq:general_solution}) can be rewritten in terms of
the GN coordinate defined with respect to the inflationary brane 
as \cite{Gor}:
\begin{eqnarray}
h(\tau,z) &=& 
\int_0^{\infty} dm \,\left\{\eta \sinh (y/\ell)\right\}^2 
\left\{ \tilde{h}_1(m) \,H_2^{(1)}\left(m \eta \sinh(y/\ell)\right)
\,\, e^{-i\,\omega \,\eta \cosh (y/\ell)} \right.
\nonumber\\
&& \left. ~~~~~~~~~~ 
+ \tilde{h}_2(m) \,H_2^{(2)}\left(m \eta \sinh (y/\ell)\right) 
\,\, e^{i\,\omega \,\eta \cosh (y/\ell)} \right\}. 
\label{eq:desitter} 
\end{eqnarray}
Comparing (\ref{eq:inflation_zero_mode}) with
(\ref{eq:desitter}), we see that the zero-mode solution given in the
inflationary epoch cannot be simply expressed in terms of the zero-mode 
solution in the Poincar\'e coordinates, indicating that  
there should be mixtures of KK-modes to express the zero-mode 
solution in the inflationary epoch. Nevertheless, 
in the long-wavelength limit $k\to0$, both the zero-mode solutions 
become constant over the time and the bulk space and they coincide with 
each other. Since we are specifically concerned with the evolution of 
long-wavelength GWs after inflation, the constant mode, i.e., 
$h=\text{const.}$ and $dh/d\tau=0$, seems a natural and a physically 
plausible initial condition for our numerical calculation in the 
Poincar\'e coordinate.

However, a subtle point arises when we consider the evolution of GWs 
in the radiation dominated epoch. In this case, 
the mode decomposition becomes generally impossible in GN coordinates 
due to the non-separable form of the background metric    
(\ref{eq: back_metric_GN})--(\ref{eq:lapse_func}),  
and the constant mode is not necessarily a solution. 
From the viewpoint of the Poincar\'e coordinate in 
the AdS${}_5$ bulk, the mixture of KK-modes could be 
significant in the high-energy regime of the universe 
and this is even true in the long-wavelength GWs. 
Indeed, even in the low-energy regime ($\rho/\lambda\ll1$), 
the mixture of KK-modes has been shown to be essential for the 
recovery of the standard four-dimensional result \cite{KK}.

Hence, the constancy of the GW amplitudes after inflation 
cannot be always guaranteed even on super-horizon scales.  
Depending on the choice of the bulk coordinate, the mode 
$h(t_0,z)=\text{const.}$ 
may not be a good approximation to the initial condition for numerical 
calculation if one tries to impose the initial condition in the radiation 
dominated epoch. In order to clarify these subtleties, 
the validity of the initial condition $h=\text{const.}$, 
must be checked. This point will be carefully discussed 
in section \ref{subsec:superhorizon}.

\subsection{Numerical simulation}
\label{subsec:numerical_simul}

To solve the wave equation (\ref{eq: wave}) numerically, 
one problem is that the computational domain should be finite. 
We must introduce an {\it artificial} cutoff (regulator) boundary 
in the bulk at $z=z_{\rm reg}$ and impose the boundary condition. 
Here, we impose the Neumann condition at the regulator 
boundary,  i.e., $(\partial h/\partial z)_{z=z_{\rm reg}}=0$.   
The location of the boundary is set to $z_{\rm reg} = 30$--$100\ell$, 
which is far enough away from the physical brane to avoid 
artificial suppression of light KK-modes. 
We checked that the amplitude of GWs on the brane is fairly 
insensitive to the location of regulator boundaries. 
Further, we stop the numerical calculations before the influence of the 
boundary condition at
$z=z_{\rm reg}$ can reach the physical brane $z_{\rm b}$. 
With these treatments, all the results presented in Section 
\ref{sec: results} are free from the  effect of regulator boundary.

The numerical calculations of wave equation (\ref{eq: wave}) were 
carried out by employing the Pseudo spectral method \cite{Can}. 
To be precise, we adopt a Tchebychev collocation method
with Gauss-Lobatto collocation points. 
To implement this, instead of using the Poincar\'e coordinates 
$(\tau, z)$ directly, we use the following new coordinates $(t,\xi)$:  
\begin{equation}
\tau=T(t),\qquad
z=\frac{1}{2}\left[\,\left\{z_{\rm reg}-z_{\rm b}(t)\right\}\xi + 
\left\{z_{\rm reg}+z_{\rm b}(t)\right\} \,\, \right]
\end{equation}
so that the locations of both the physical and the regulator branes 
are kept fixed. Adopting this coordinate system, 
the perturbed quantity $h(t,\xi)$ is first transformed 
into the Tchebychev space 
through the relation, $h(t,\xi)= \sum^N_{n=0} h_n(t)T_n(\xi)$   
in a finite and a compact domain, $-1\leq \xi \leq 1$. 
Here, the functions $T_n(\xi)$ denote the 
Tchebychev polynomials, defined by $T_n(\xi)\equiv \cos(n\cos^{-1}(\xi))$. 
We then discretize the $\xi$-axis to the $N+1$ points (collocation points) 
using the inhomogeneous grid $\xi_n=\cos(n\pi/N)$. With this grid, 
fast Fourier transformation can be applied to 
perform the transformation between the amplitude $h(t,\xi)$ and 
the coefficients $h_n(t)$. In this letter, we specifically set 
the collocation point as $N=512$ or $1024$. 
The partial differential equation (\ref{eq: wave}) is now reduced 
to a set of ordinary differential equations for the coefficients 
$h_n(t)$. Hence,  
one can obtain the time evolution of $h_n(t)$ by simply adopting the 
Predictor-Corrector method based on the Adams-Bashforth-Moulton scheme.

\section{Results}
\label{sec: results}

Given the initial condition $h(t_0,\xi)=\text{const.}$, 
the remaining free parameters in our numerical simulation 
are the wave number $k$ and the initial time $t_0$. 
For convenience, we set the wave number 
$k$ to $k=a_*H_*$, that is, the GW 
just crosses the Hubble horizon at the time $t=t_*$ 
(see Eq.(\ref{eq: scale_energy})). 
Also, we introduce the quantity $s_0$ given by  
$s_0 \equiv  a(t_0)H(t_0)/k$, 
which represents the physical scale of the long-wavelength GW  
normalized by the Hubble horizon at an initial time $t_0$. 
Thus, the free parameters may be represented 
by the dimensionless energy at the horizon-crossing time, 
$\epsilon_*\equiv\epsilon(t_*)$ 
and the normalized wavelength at initial time, $s_0$. 
In the following, the numerical results are presented for various 
choices of the parameters $(s_0, \epsilon_*)$.

\subsection{Sensitivity to the initial condition}
\label{subsec:superhorizon}

\begin{figure}[ht]
 \centering
 \includegraphics[width=14cm]{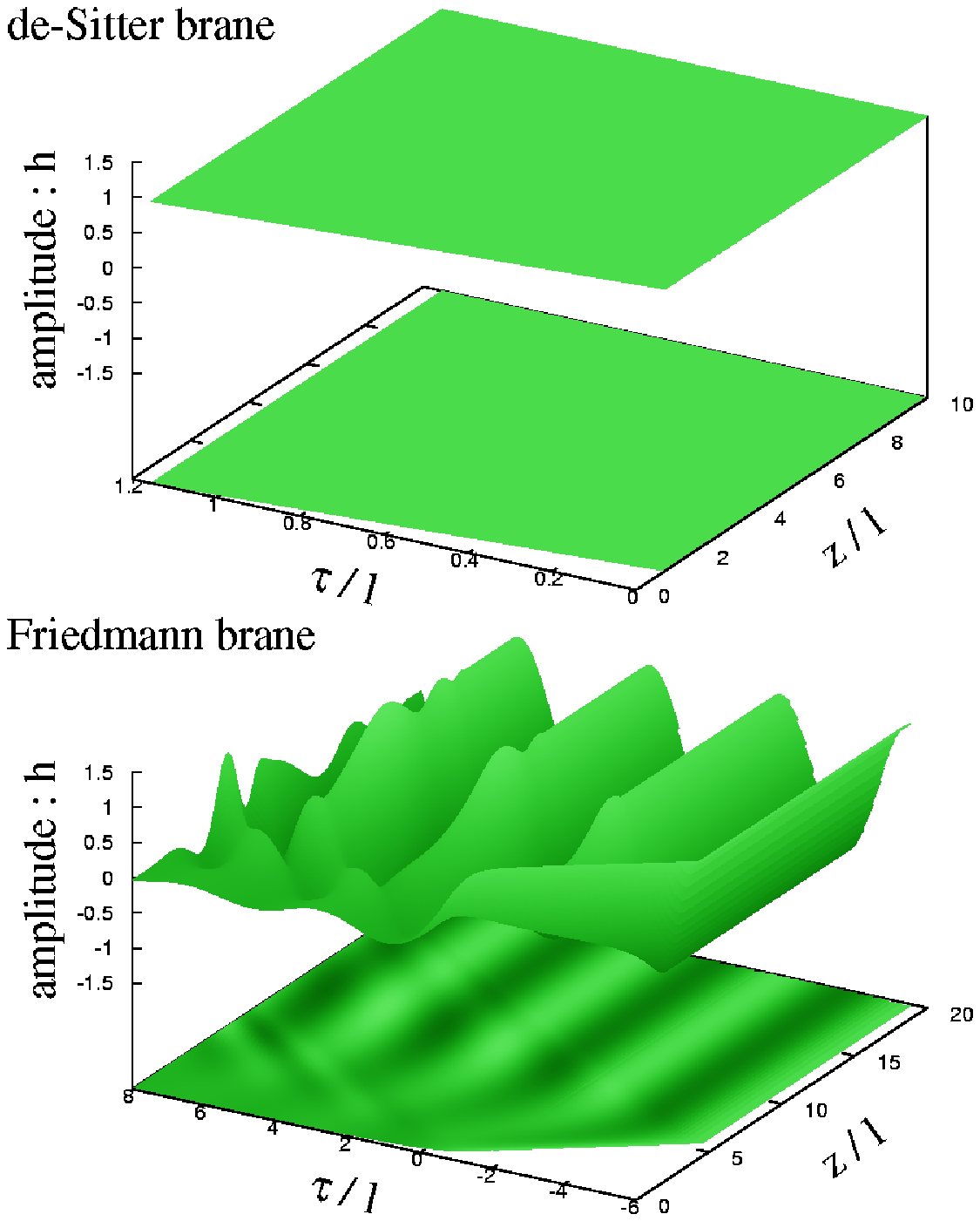}
 \caption{The evolution of a GW in the bulk. The upper panel shows the
 case of a de Sitter brane, while the lower panel is the case of a
 Friedmann brane. In both panels, we set the comoving wave number to 
 $k=\sqrt{3}/\ell$ or $\epsilon_*=1.0$ with $s_0=100$.  
 \label{fig: 5D_behavior}} 
\end{figure}

Let us first consider the initial condition after inflation 
and check the validity of the assumption $h(t_0,\xi)=\text{const.}$ 
for long-wavelength GWs. For this purpose, 
we plot the time evolution of GWs in the Poincar\'e coordinate system 
in Figure \ref{fig: 5D_behavior}. 
The upper panel shows the case of the de Sitter brane by setting the 
equation of state on the brane to $w=-1$. The lower panel shows 
the case of the radiation-dominated Friedmann brane $(w=1/3)$.   
In both panels, we set the comoving wave number to 
$k=\sqrt{3}/\ell$ or $\epsilon_*=1.0$ with $s_0=100$.

In the upper panel of Figure \ref{fig: 5D_behavior}, 
the universe on the brane experiences  
accelerated cosmic expansion and the wavelength of GWs becomes longer
than the Hubble horizon. The resultant GW amplitude 
remains constant not only on the brane but also in the bulk. 
On the other hand, in the case of the Friedmann brane ({\it lower}), the
wavelength of GW becomes shorter than the Hubble horizon at $\tau
\approx -1$. In the bulk, a complicated oscillatory behavior of
perturbations was found in the region that is causally connected to the
physical brane. This indicates that the excitation of KK-modes
occurs even if the wavelength of GWs is still outside the Hubble horizon. 

These results are somewhat surprising from the viewpoint of the AdS${}_5$ 
bulk, because the different behaviors simply arise from the difference
in the motion of the brane. While the trajectory of the brane is described 
by a straight line ($d^2z_{\rm b}/d\tau^2=0$) in the case of the de 
Sitter brane, the trajectory of the Friedmann brane describes an arc with 
a non-zero curvature $d^2z_{\rm b}/d\tau^2<0$. The situation might be very 
similar to the moving mirror problem in an electromagnetic field 
(e.g., Chap. 4.4 of Ref. \cite{BD}),  where the acceleration  or 
deceleration of the mirror yields the creation of photons due to 
vacuum polarization. In our present case, 
the excitation of the KK-modes arises due to the deceleration of the 
moving brane. 

The results in Figure \ref{fig: 5D_behavior} suggest that 
the initial condition $h(t_0,\xi)=\text{const.}$ 
may be validated if we set the initial condition just after the 
end of inflation. However, the constancy of the long-wavelength 
mode would not be guaranteed in the case of the radiation 
dominated epoch, as discussed in section \ref{subsec:Initial_condition}. 
This implies that the choice of the initial time 
$t_0$ is crucial when setting the initial condition at the radiation 
dominant epoch. Thus, for quantitative investigation of  
the GWs generated during inflation, the sensitivity to the choice of 
the initial time should be examined.

\begin{figure}
 \centering
 { 
\includegraphics[width=6.7cm]{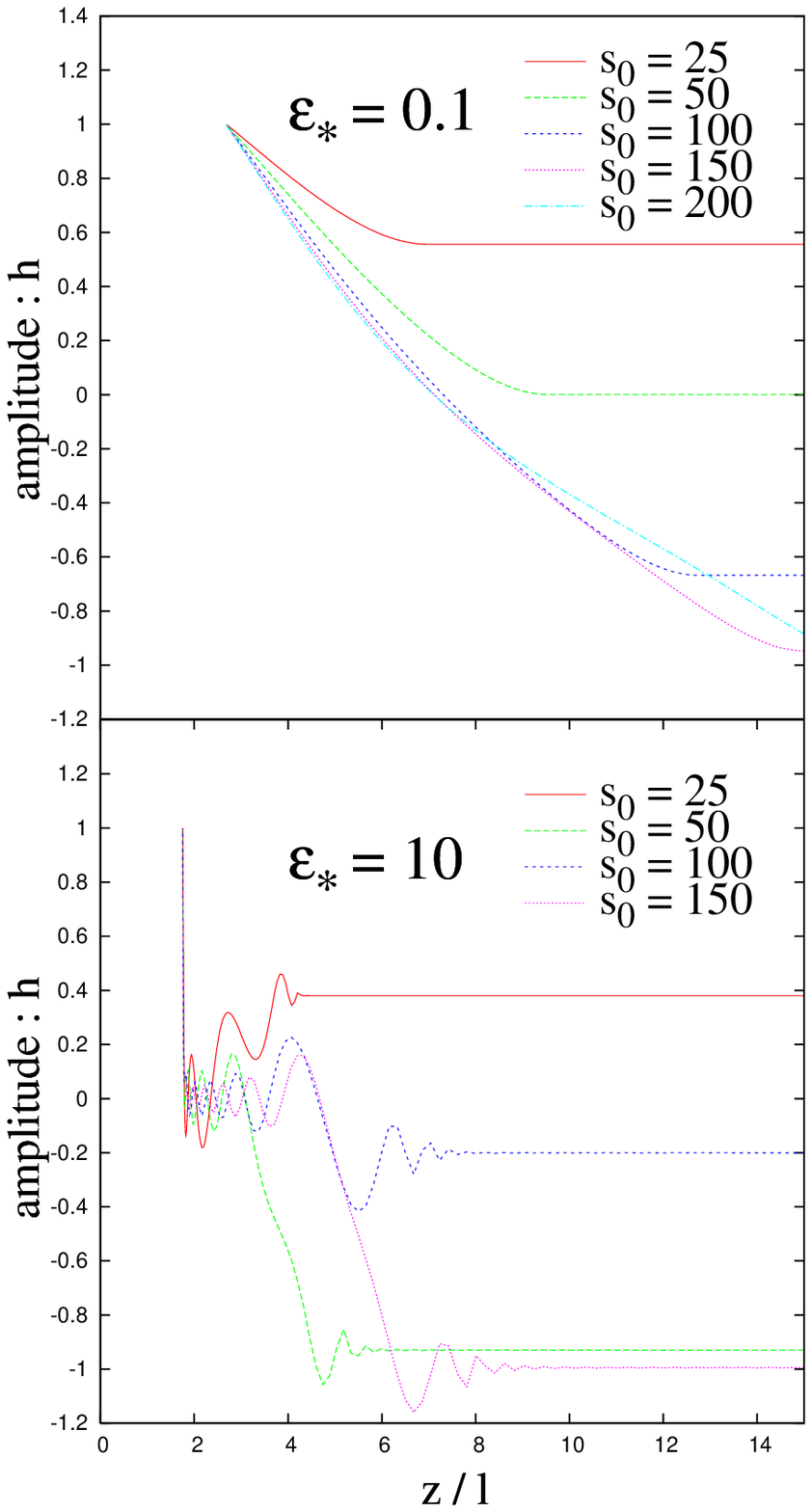}
\includegraphics[width=7cm]{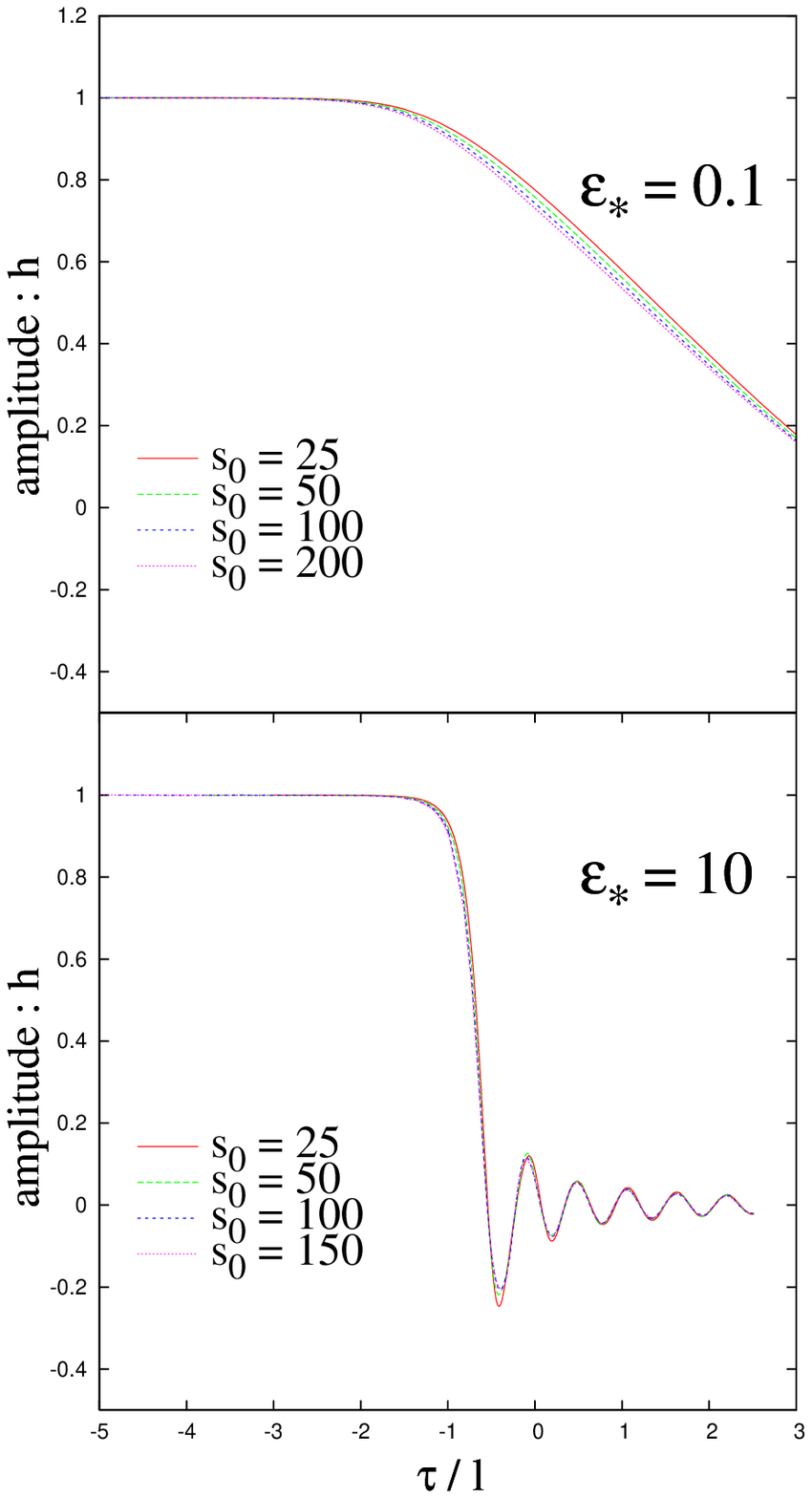}
 }
\caption{{\it Left}: 
Snapshots of the GW amplitudes in the bulk for various choices
 of initial time. The snapshots were taken when the 
 wavelength of GWs becomes five times longer than the Hubble horizon, 
i.e., $a_0H/k=5$. {\it Right}: Evolved results of GWs projected on the brane 
starting with the various initial times.
 \label{fig: sensitivity_to_t0}}
\end{figure}

In Figure \ref{fig: sensitivity_to_t0}, the dependence of 
the evolution of GWs on the initial time is plotted by varying the 
parameter $s_0$. Left panels show the snapshots of the amplitude $h(\tau,z)$ 
in the bulk when the wavelength of GWs just becomes five times 
larger than the Hubble horizon, i.e., $a_0H/k=5$.  On the 
other hand, right panels show the time evolution of GWs 
projected on the brane. Clearly, in the bulk, the amplitude of 
GWs is very sensitive to the choice of the parameter $s_0$, or 
equivalently, the initial time $t_0$. The resultant wave form away from 
the physical brane does not show any convergence
even in the low-energy case $(\epsilon_*=0.1)$.  
By contrast, on the brane, the GW amplitudes 
tend to converge if we set the initial time $t_0$ early enough 
(or set $s_0$ large enough). 

Although we do not fully understand the reason for this 
convergence, as far as the GWs on the brane are 
concerned, the evolution of GW amplitudes becomes insensitive to the choice 
of the initial time when setting the parameter $s_0$ large enough, 
$s_0\gtrsim50$, for instance. 

\subsection{Influence of high-energy effects on spectrum of 
gravitational wave background}

Having validated the setup of the initial conditions, 
we now attempt to clarify the high-energy effects of the GWs and 
evaluate the spectrum of the SBGW on the brane. 
To quantify these, we wish to discriminate the 
influence of KK-mode excitation in the bulk from 
the non-standard cosmological expansion caused by the 
$\rho^2$-term in the Friedmann equation (\ref{eq: Friedmann}).  
For this purpose, we introduce the reference wave $h_{\rm ref}$,  
which is a solution of the four-dimensional wave equation just replacing 
the scale factor and the Hubble parameter defined in the 
standard Friedmann equation with those defined in 
the modified Friedmann equation (Eqs. (\ref{eq: Friedmann}) and 
(\ref{eq: scale_energy})):
\begin{equation}
  \ddot{h}_{\rm ref}+3H\dot{h}_{\rm ref}+\left(\frac{k}{a_0}\right)^2
h_{\rm ref}=0.
  \label{eq: reference}
\end{equation}
Comparing the numerical simulations with the
solution of the wave equation (\ref{eq: reference}), 
the effect of the excitation of KK-modes can be quantified. 

\begin{figure}
 \centering
 { \includegraphics[width=10cm]{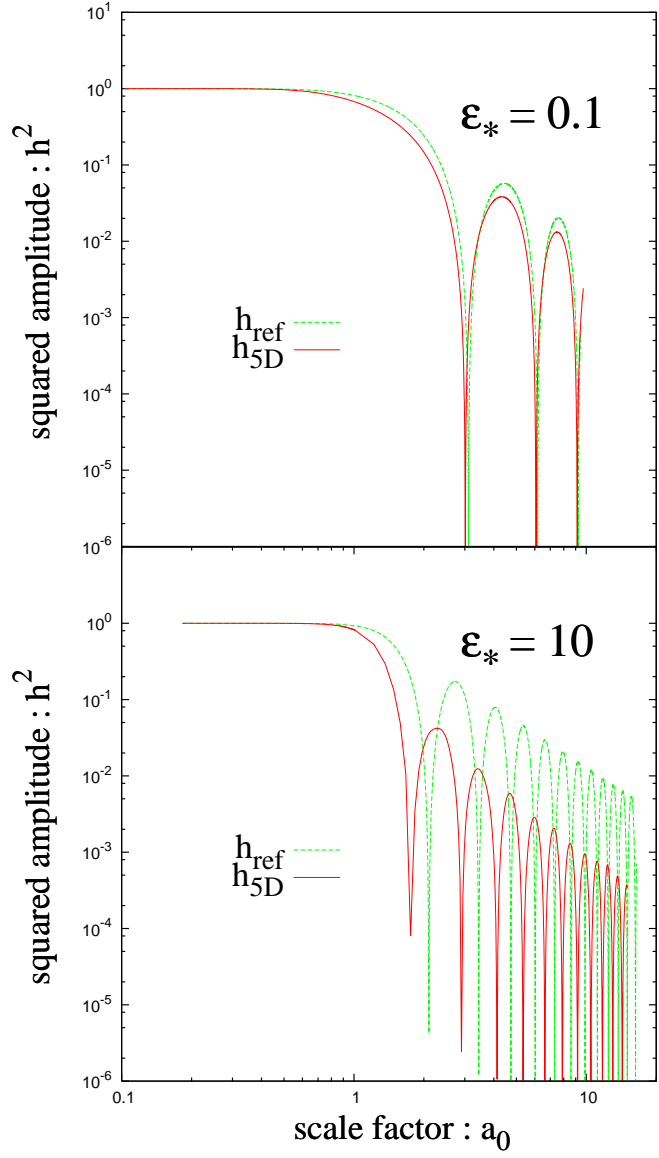}}
 \caption{Squared amplitude of GWs on the brane in low-energy ({\it
 upper}) and the high-enery ({\it lower}) regimes. In both panels, solid 
 lines represent the numerical solutions of wave equation (\ref{eq:
 wave}). The dashed lines are the amplitudes of reference wave $h_{\rm ref}$
 obtained from equation (\ref{eq: reference}). 
\label{fig: brane_behavior}}
\end{figure}

Figure \ref{fig: brane_behavior} shows the squared amplitude of 
the GWs, $h_{\rm 5D}^2$ and $h_{\rm ref}^2$ as functions of the scale
factor $a_0$.
The upper panel shows the low-energy case ($\epsilon_*=0.1$), 
while the  lower panel depicts the result in the high-energy regime 
($\epsilon_*=10$). In both panels, the horizon re-entry time is 
set to $a_0=1$. As we increase the 
energy scale at horizon crossing time, the GW amplitude 
$h_{\rm 5D}$ becomes significantly reduced compared to the 
reference wave, $h_{\rm ref}$. Since the late-time evolution of 
GWs simply scales as $h\propto 1/a_0$ in both $h_{\rm 5D}$ and $h_{\rm ref}$, 
the results may be interpreted 
as the excitation of KK-modes during horizon re-entry, 
which are caused by an escape of five-dimensional graviton
from the brane to the bulk. 
Note that the normalized energy density at  
horizon re-entry time, $\epsilon_*$ is related to the observed proper 
frequency $2\pi\,f=k/a_0(t_{\rm today})$ as 
\begin{equation}
\frac{f}{f_{\text{crit}}}=\left(\frac{\sqrt{2}-1}{\epsilon_*}\right)^{1/4}(\ell 
H_*)^{1/2} 
= \left(\frac{\sqrt{2}-1}{\epsilon_*}\right)^{1/4}(\epsilon_*^2+2\epsilon_*)^{1/4},
  \label{eq: relation_with_eps}
\end{equation}
where the critical frequency $f_{\text{crit}}$ is defined by 
$\ell H_*=1$ or $\epsilon_*=\sqrt{2}-1$, which typically yields 
$f_{\rm crit}=0.2$mHz$\,(0.1{\rm mm}/\ell)^{1/2}$ \cite{Hogan}. 
Thus, one expects from Figure \ref{fig: brane_behavior} 
that the deviation from the 
standard four-dimensional prediction for the spectrum of 
SBGW becomes more prominent above 
the critical frequency, $f>f_{\rm crit}$.

\begin{figure}
 \centering
 \includegraphics[width=12cm]{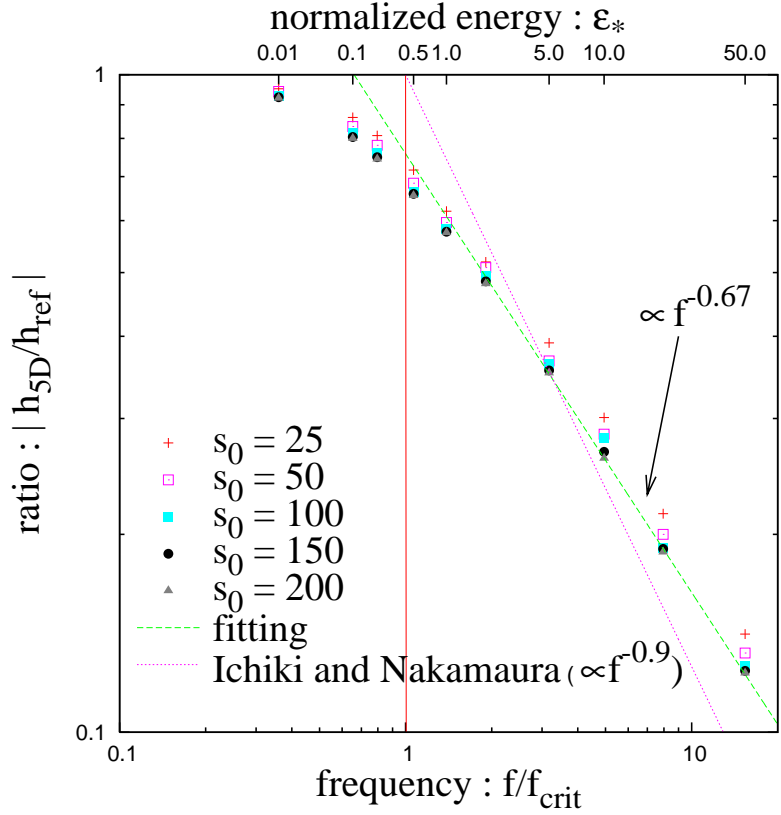}
 \caption{Frequency dependence of the ratio of amplitudes 
 $|h_{\rm 5D}/h_{\rm ref}|$ between the numerical simulation of wave 
 equation (\ref{eq: wave}) and the reference wave (\ref{eq:
 reference}). The vertical solid line represents the critical
 frequency. The dashed line indicates the fitting result (\ref{eq:
 fitting}), where fitting was examined using the data with $s_0=200$  at the 
asymptotic region $\epsilon_* \geq 5$. The dotted line shows the result by
 Ichiki and Nakamura \cite{IN2}. \label{fig: ampratio}}
\end{figure}

In order to estimate the influence of KK-mode excitation on the 
spectrum of SBGW, the frequency dependence of the GW amplitudes 
is examined based on the results in Figure \ref{fig: brane_behavior}. 
Figure \ref{fig: ampratio} shows the ratio of the amplitudes  
$|h_{\rm 5D}/h_{\rm ref}|$ as a result of the KK-mode excitation. 
The ratio is evaluated at the low-energy regime long after the 
horizon crossing time and is plotted as a function of the 
frequency $f/f_{\rm crit}$. As a reference, we also show the 
convergence properties of the ratio $|h_{\rm 5D}/h_{\rm ref}|$ 
by varying the initial time; $s_0=25,\,50,\,100,\,150$ and $200$. 
Clearly, the ratio $|h_{\rm 5D}/h_{\rm ref}|$ monotonically decreases 
with the frequency and the suppression of amplitude $h_{\rm 5D}$ 
becomes significant above the critical frequency $f_{\rm crit}$ 
(vertical solid line).  Using the data points in  
the asymptotic region $\epsilon_*\geq 5$ (or $H_*\ell \geq \sqrt{35}$), 
we try to fit the ratio of amplitudes 
with $s_0=200$ to a power-law function. 
The result based on the least-square method becomes
\begin{equation}
  \left|\frac{h_{\rm 5D}}{h_{\rm ref}}\right| = 
\alpha\left(\frac{f}{f_{\text{crit}}}\right)^{-\beta}
  \label{eq: fitting}
\end{equation}
with $\alpha=0.76\pm0.01$ and $\beta=0.67\pm0.01$, 
which is shown as a dashed line in Figure \ref{fig:
ampratio}\footnote{The uncertainty of the fitting parameters is
estimated assuming that each data point follows a standard normal
distribution.}.

The power-law fit (\ref{eq: fitting}) can be immediately 
translated to the spectrum of SBGW as follows. To evaluate the spectrum, 
we conventionally use the density parameter of SBGW 
$\omegagw(f)$ defined by \cite{Mag} :
\begin{equation}
  \omegagw(f) \equiv \frac{1}{\rho_c}
\frac{d\rho_{\rm\scriptscriptstyle GW}}{d\log f} \propto f^2h^2, 
\label{eq:Omega_gw}
\end{equation}
where $\rho_{\rm\scriptscriptstyle GW}$ is the energy density of SBGW  
and $\rho_c$ is the critical density. 
Above the critical frequency $f>f_{\rm crit}$, the cosmological 
expansion at horizon re-entry time is dominated by the 
$\rho^2$-term in the Friedmann equation (\ref{eq: Friedmann}) 
and we have $H_*\propto a_*^{-4}$. 
For the reference wave $h_{\rm ref}$, this gives 
$h_{\rm ref}\propto a_*/a_0 \propto f^{-1/3}$, because of the relation 
$2\pi f\propto k\,\propto\,a_*H_*$. Thus, substituting this 
into the expression (\ref{eq:Omega_gw}), the blue spectrum 
$\omegagw\,\propto\,f^{4/3}$ is obtained if we neglect the 
excitation of KK-modes. Consequently, from (\ref{eq:
fitting}), if we combine the effects of the KK-mode excitation,
the SBGW spectrum becomes nearly flat, i.e., 
\begin{equation}
\omegagw\,\propto\,f^{0}, 
\end{equation}
above the critical frequency. 
This result seems almost indistinguishable from the standard four-dimensional
prediction and slightly contradicts with the result by 
Ichiki and Nakamura \cite{IN2}, who predict a relatively steeper  
spetrum, $\omegagw\,\propto\,f^{-0.46}$ from the numerical simulation
based on a null coordinate system. At present, 
the reason for this discrepancy is unclear, however, 
it might be ascribed to the differences of initial conditions arising 
from the different choices of the bulk coordinates. 
Although it is premature to discuss the detectability of the SBGW,
the significance of KK-mode excitations should play an important clue to
probe the signature of extra-dimensions.


\section{Summary and Discussion}
\label{sec: summary}

We numerically studied the behavior of GWs in the high-energy regime of 
the universe in the Randall-Sundrum brane-world model.  
In contrast to the previous study in the GN coordinates, 
we used the Poincar\'e coordinate system in which the
brane is moving in the AdS${}_5$ background. 
We numerically investigated the evolution of GWs in the radiation
dominated universe.

We have confirmed that the constant mode, $h(z,\tau)=$const.,   
in the Poincar\'e coordinates just after inflation is a plausible 
initial condition from inflation. In the radiation dominated epoch, 
however, the constant mode does not hold in the bulk. Thus, in
principle, we have to start our numerical calculations just after 
inflation. Fortunately, however, the resultant amplitude of 
the GWs on the brane becomes insensitive 
to the initial time $t_0$ of our numerical simulation 
if we set $t_0$ sufficiently early. 
In other words, our numerical result is insensitive to the  
transition time from inflation to the radiation dominated 
epoch if the perturbation re-enters the horizon long enough 
after the transition. 


Based on these discussions,  we turn to focus on the frequency dependence
of the high-energy effects in order to predict the spectrum of 
SBGW. We found that the excitation of KK-modes compensates the 
effect of non-standard cosmological expansion. As a result, 
the density parameter of SBGW scales as $\omegagw\,\propto\,f^{0}$ 
above the critical frequency
$f_{\text{crit}}=0.2 \text{mHz}\,(0.1\text{mm}/\ell)^{1/2}$.  
The result seems indistinguishable from the prediction in the 
standard four-dimensional theory and slightly contradicts with 
the result by Ichiki and Nakamura \cite{IN2}. 
The most striking result of our work is that the behavior of the
GWs in the bulk sensitively depends on the transition time. 
This feature implies that the constant initial condition in the Poincar\'e
coordinate does not agree with the constant initial condition on the
null hypersurface even on super-horizon scales. In this sense, the
result obtained by Ichiki and Nakamura \cite{IN2}  may not necessarily
agree with our results.  In addition, the present numerical analysis is
restricted to the frequency  range $0.3\lesssim f/f_{\rm crit}\lesssim
15$. It is thus premature to discuss the detectability of the SBGW
by extrapolating our numerical results  to the high-frequency bands
observed via future GW detectors.  
Nevertheless, the significance of KK-mode excitations above the critical
frequency holds the clue to probe the signature of
extra-dimensions and/or to constrain on brane-world models. In this
respect, a more quantitative and precise prediction for the spectrum of
SBGW should deserve a further investigation.

Recently, Kobayashi and Tanaka \cite{KT} analytically estimated the effect 
of KK-mode excitation at low-energy scales.  
According to their result, the amplitude of the GWs on the brane depends 
on the transition time as well as the energy scale $\epsilon_*$. 
This would not be a contradiction with the present numerical result
because we set the initial condition in the high-energy regime $H \ell
>1$, where the low-energy approximation used by the authors does not work. 
Thus, in order to understand our result, we need an analytical study 
of the KK-mode excitation in the high-energy regime. 
This is a challenging future work. 

\begin{ack}
We would like to thank Roy Maartens for reading our  
manuscript carefully and for his hospitality during
TH's visit to Portsmouth. We would like to thank  
Kiyotomo Ichiki for discussions and comments. KK and TH 
acknowledge the support from the Japan Society for Promotion of 
Science(JSPS) Research Fellowships. AT is supported 
by a Grant-in-Aid for Scientific Research from the JSPS(No.14740157). 
\end{ack}



\clearpage
\end{document}